\documentclass[twocolumn,showpacs,amsmath,pre]{revtex4}

\usepackage{graphicx}
\usepackage{amsmath}
\usepackage{amssymb}
\usepackage{natbib}

\begin{document}

\title{Tuning Jammed Frictionless Disk Packings from Isostatic to Hyperstatic}

\author{Carl F. Schreck$^{1}$}
\author{Corey S. O'Hern$^{2,1}$}
\author{Leonardo E. Silbert$^{3}$} 

\affiliation{$^{1}$Department of Physics, Yale University, New Haven,
  Connecticut 06520-8120, USA}

\affiliation{$^{2}$Department of Mechanical Engineering, Yale University, New
  Haven, Connecticut 06520-8260, USA}

\affiliation{$^{3}$Department of Physics, Southern Illinois University,
  Carbondale, Illinois 62901, USA}

\begin{abstract}
We perform extensive computational studies of two-dimensional static
bidisperse disk packings using two distinct packing-generation
protocols.  The first involves thermally quenching equilibrated liquid
configurations to zero temperature over a range of thermal quench
rates $r$ and initial packing fractions followed by compression and
decompression in small steps to reach packing fractions $\phi_J$ at
jamming onset.  For the second, we seed the system with initial
configurations that promote micro- and macrophase-separated packings
followed by compression and decompression to $\phi_J$.  Using these
protocols, we generate more than $10^4$ static packings over a wide
range of packing fraction, contact number, and compositional and
positional order.  We find that amorphous, {\it isostatic} packings
exist over a finite range of packing fractions from $\phi_{\rm min}
\le \phi_J \le \phi_{\rm max}$ in the large-system limit, with
$\phi_{\rm max} \approx 0.853$.  In agreement with previous
calculations, we obtain $\phi_{\rm min} \approx 0.84$ for $r > r^*$,
where $r^*$ is the rate above which $\phi_J$ is insensitive to
rate. We further compare the structural and mechanical properties of
isostatic versus hyperstatic packings.  The structural
characterizations include the contact number, bond orientational
order, and mixing ratios of the large and small particles.  We find
that the isostatic packings are positionally and compositionally
disordered, whereas bond-orientational and compositional order
increase with contact number for hyperstatic packings. In addition, we
calculate the static shear modulus and normal mode frequencies of the
static packings to understand the extent to which the mechanical
properties of amorphous, isostatic packings are different from
partially ordered packings.  We find that the mechanical properties of
the packings change {\it continuously} as the contact number increases
from isostatic to hyperstatic.
\end{abstract}

\pacs{
83.80.Fg
61.43.-j
62.20.-x
64.75.Gh
}

\maketitle

\section{Introduction}
\label{intro}

The ability to enumerate and classify all of the mechanically stable
(MS) packings of frictionless particles is important for understanding
glass transitions~\cite{stillinger0} in atomic, molecular, and
colloidal systems, and the structural and mechanical properties of
particulate materials such as granular media, foams, and emulsions.
For example, if all MS packings in a given system are known, one can
measure accurately the frequency with which each MS packing occurs,
and determine how the packing frequencies and materials properties
depend on the preparation history~\cite{ning,gao}.  Further, MS packing
frequencies are important for identifying the appropriate statistical
mechanical ensemble for weakly perturbed granular
materials~\cite{song}.  However, since the number of MS packings grows
exponentially with the number of particles~\cite{stillinger}, exact
enumeration of static packings is prohibitive for even modest system
sizes~\cite{shattuck}.  Thus, one of the most important outstanding
questions in the area of disordered particulate materials is
determining how the packing-generation protocol influences the
distribution of MS packings and their structural and mechanical
properties.

Previous work has suggested that the positional order of MS packings
of frictionless spheres increases monotonically with packing fraction
and contact number in dense packings~\cite{torquato,kansal}.  However,
the MS packings in these previous studies were created using
monodisperse systems, which are prone to
crystallization~\cite{torquato3}, and prepared using the
Lubachevsky-Stillinger compression algorithm~\cite{ls}, which is a
thermalized packing-generation protocol.  In addition, these prior
studies did not distinguish the distribution of isostatic MS packings
(in which the number of degrees of freedom matches the number of
constraints~\cite{witten}) from the distribution of hyperstatic
packings (with more contacts than degrees of freedom).  Later work
characterized bidisperse systems, which are less prone to
crystallization, but focused on microphase-separated states, not
amorphous, isostatic packings~\cite{donev}.  However, recent studies
on systems composed of 3D monodisperse, frictionless, spherical
particles have pointed out that amorphous, isostatic packings can
exist over a finite range of packing fraction in the large-system
limit, with no correlation between positional order and packing
fraction~\cite{berthier2,chaudhuri}.  Moreover,
simulations~\cite{vagberg} and experiments ~\cite{lechenault} on
two-dimensional systems also suggest a finite range of jamming onsets
rather than a single packing fraction in the large system limit.

Further, the body of work on jammed particulate systems has emphasized
the concept of point J, {\it i.e.} that there is a single packing
fraction at which jamming occurs in the large system
limit~\cite{longJ,kamien}.  Since amorphous, isostatic packings can
exist over a finite range of packing fractions, the onset of jamming
should not be classified as a point in the jamming phase
diagram, but rather as a region of finite extent.  It has
also been argued that the wide distribution of packing fractions at
which the onset of jamming occurs in small periodic
systems~\cite{longJ} is related to the finite range of packing
fractions over which amorphous, isostatic packings occur in the large
system limit~\cite{bible}.  However, it has not been proved that these
two effects are directly connected.   
  
A number of overarching questions related to the connection between
positional order, isostaticity, and material properties of static
packings remain open. For example, can isostatic or nearly isostatic
packings possess significant positional order and if so, what are the
fundamental differences in the normal modes and mechanical
properties between those that do and do not possess significant positional
order?  This question is particularly important since recent studies
have emphasized that {\it amorphous}, isostatic packings possess an
excess of low-frequency normal modes~\cite{liu,mao} over that
for harmonic, ordered solids.

In addition, previous work has drawn a strong contrast between
amorphous packings and configurations with crystalline
order~\cite{torquato2}.  However, how different are the structural and
mechanical properties of amorphous versus partially ordered
particulate systems?  For example, it is possible that the amorphous
regions in the interstices between ordered domains in partially
crystalline materials dominate the structural and mechanical
properties, in which case their properties would be similar to
amorphous packings.  At the very least, one would assume that there is
not a strong difference between the mechanical properties of isostatic 
and only slightly hyperstatic packings that possess significant positional 
order. 

In this article, we describe extensive computer simulations of
collections of frictionless, bidisperse disks with short-range
repulsive interactions to address two important, open questions:
1. What is the range of packing fractions over which amorphous,
isostatic static packings occur with similar structural and mechanical
properties, and 2. How do the structural and mechanical properties of
static packings change with the deviation in the contact number at
jamming onset from the isostatic value, $z_J - z_{\rm
iso}$~\cite{foot2}?  Using two distinct packing-generation protocols,
we construct scatter plots for more than $10^4$ static packings
characterized by the contact number, packing fraction, measures of
positional order, and mechanical properties.  The first protocol
involves thermally quenching equilibrated liquid configurations to
zero temperature over a range of thermal quench rates $r$ followed by
compression and decompression in small steps to reach packing
fractions $\phi_J$ at jamming onset.  For the second, we seed the
system with initial configurations that promote micro- and
macrophase-separated packings followed by compression and
decompression to $\phi_J$.

Our main results are fourfold: 1. Isostatic, amorphous packings exist
over a finite range of packing fraction from $\phi_{\rm min}$ to
$\phi_{\rm max}$ in the large system limit, with similar structural
and mechanical properties.  2. In agreement with previous
calculations, we obtain $\phi_{\rm min} \approx 0.84$ for $r > r^*$,
where $r^*$ is the rate above which $\phi_J$ is insensitive to
rate. In contrast, $\phi_{\rm max}$ depends sensitively on quench
rate, system size, and boundary conditions. 3) The amorphous,
isostatic packings coexist with an abundance of hyperstatic,
microphase- and macrophase-separated packings. 4) When considering the
full ensemble of static frictionless packings, the packings possess
structural and mechanical properties that span a continuous range from
amorphous to partially ordered to ordered in contrast to the results
and interpretations of recent studies~\cite{makse,radin}.
 
The remainder of the manuscript will be organized as follows. In
Sec.~\ref{protocol}, we describe the computational system we consider
and the two protocols we employ to generate static frictionless disk
packings.  In Sec.~\ref{characterize}, we present our results, which
include characterizations of the structural (packing fraction, contact
number, and several order parameters to detect positional and
compositional order) and mechanical (shear modulus and eigenvalues of
the dynamical matrix~\cite{gao}) properties of more than $10^4$ static
packings and comparisons of these properties for isostatic and
hyperstatic configurations.  Finally, in Sec.~\ref{conclusions}, we
provide our conclusions and promising future research directions.
   
\section{Packing-Generation Protocols}
\label{protocol}

We focus on well-characterized two-dimensional systems composed of $N$
bidisperse disks ($50$-$50$ by number), each of mass $m$, with
diameter ratio $d=\sigma_l/\sigma_s=1.4$~\cite{harrowell,longJ,donev},
within square, periodic simulation cells with side length $L$.  We
consider frictionless particles that interact through the
finite-range, purely repulsive spring potential.  The total potential
energy per particle is given by
\begin{equation}
\label{interaction}
V = \frac{\epsilon}{2N} \sum_{i>j} \left( 1 - \frac{r_{ij}}{\sigma_{ij}}
\right)^{2} \Theta \left( 1 - \frac{r_{ij}}{\sigma_{ij}} \right),
\end{equation}
where $r_{ij}$ is the center-to-center separation between disks $i$
and $j$, $\epsilon$ is the characteristic energy scale of the
interaction, $\Theta(x)$ is the Heaviside function, and $\sigma_{ij} =
(\sigma_{i}+\sigma_{j})/2$ is the average diameter. We simulated a
range of system sizes from $N=256$ to $8192$ particles to assess
finite size effects.  Energy, length, and time scales are measured in
units of $\epsilon$, $\sigma_s$, and $\sigma_s \sqrt{m/\epsilon}$,
respectively.

The packing fraction $\phi_J$ at which jamming occurs and the
structural and mechanical properties of static packings can depend
strongly on the packing-generation protocol employed.  Our goal is to
generate static frictionless MS packings that span the range of contact numbers
from the isostatic value $z_{\rm iso} = 4$ to the hexagonal crystal
value $z_{\rm xtal} = 6$ and the range of positional order from
amorphous to phase-separated and from partially crystalline to
crystalline states.  To accomplish this, we investigate two distinct classes of
packing-generation protocols: 1) thermal quenching from liquid initial
conditions coupled with compression and decompression steps, which
typically generates amorphous configurations and 2) compression and
decompression steps from initial conditions that promote micro- or
macrophase separation~\cite{phase}.

{\it Protocol 1: Thermal quenching from liquid initial conditions}
In this algorithm, we prepare equilibrated, liquid configurations at
high temperature $T_0 = 10^{-3}$ and in molecular dynamics (MD)
simulations quench them to a very low final temperature $T_f=10^{-16}
\simeq 0$ at fixed packing fraction $0.8 \leq \phi_i < \phi_{\rm xtal} =
\pi/2\sqrt{3}$~\cite{foot4} over a time interval $t$ by rescaling the particle
velocities so that the kinetic temperature $T = N^{-1} \sum_i m
v_i^2/2$ obeys
\begin{equation}
T(t) = T_{0}e^{-rt}, 
\label{eq2}
\end{equation}
where $r$ is the thermal quench rate, which is varied over five orders
of magnitude $10^{-5} \le r \le 1$.  We generated $50$ equilibrated,
independent liquid configurations at $T_0$ at each $\phi_i$ by writing
out configurations every $10 \tau$, where $\tau$ is a decay time
obtained from the self-intermediate scattering function at wavenumbers
corresponding to the first peak in the structure factor~\cite{isf}.

After reaching a local potential energy minimum at each initial
packing fraction $\phi_i$ and thermal quench rate $r$, we input the
configurations into an `athermal' algorithm (`packing finder') that
searches for the nearest static packing in configuration space with
infinitesimal particle overlaps. The algorithm has been described in
detail in previous work~\cite{gao}. Briefly, we successively increase
or decrease the diameters of the grains (while maintaining the
diameter ratio $d$), with each compression or decompression step
followed by conjugate gradient minimization of $V$.  The system is
decompressed when the total potential energy per particle at a local
minimum is nonzero, {\it i.e.} there are finite particle overlaps.  If
the potential energy of the system is zero and gaps exist between
particles, the system is compressed.  The increment by which the
packing fraction is changed at each compression or decompression step
is gradually decreased.  Numerical details of the algorithm are the
same as in Ref.~\cite{gao}.  When this algorithm terminates, we obtain
a static packing defined by the particle positions $\{ {\vec r}_1,
{\vec r}_2,\ldots, {\vec r}_N\}$ and packing fraction $\phi_J$.  Since
we use an energy tolerance (per particle) $V_{\rm tol}/\epsilon =
10^{-16}$ for the termination of the energy minimization and
compression/decompression scheme in the packing finder, the positions
and packing fraction at jamming are extremely accurate with errors at
one part in $10^8$.
   
{\it Protocol 2: Compression and decompression steps from initial
conditions that promote order} We will see below in
Sec.~\ref{characterize} that Protocol $1$ produces amorphous,
isostatic packings.  Thus, we seek an algorithm that will generate
static packings with variable positional and compositional order.  To
bias the system toward micro- and macrophase-separated configurations,
we seed the packing finder with particular sets of initial conditions.
We first divided the unit cell into $s\times s$ equal-sized
partitions, where $s$ is an even integer that ranged from $2$ to $26$,
and placed approximately $N/s^2$ large or small particles in
alternating partitions to create a checkerboard-like pattern. The
particles were placed randomly in each partition.  The initial
configuration is then input into the packing finder to yield a static
packing.  In the large $s$ limit, we expect amorphous static packings,
while at intermediate and small $s$, we expect micro- and
macrophase-separated packings.  To generate static packings near
$\phi_{\rm xtal}$ we also divided the unit cell into two partitions
and placed the large (small) particles on a hexagonal lattice in a
region with area $A_L=d^2/(1+d^2)$ ($1-A_L$) and then applied the
packing finder.

\section{Structural and Mechanical Properties}
\label{characterize}

After generating static packings using the two packing-generation
protocols described above, we contrast them by calculating several
structural and mechanical properties.  The structural
characterizations include the packing fraction, contact number, and
compositional and positional order parameters.  For the packing
fraction at jamming onset, we calculate
\begin{equation}
\label{packing_fraction}
\phi_J = \frac{N\pi}{8} \left( \frac{ \sigma_s}{L}\right)^2 \left( 1 + d^2 \right)
\end{equation}
including all $N$ particles.  For the contact number at jamming, we
sum up all overlapping pairs ($r_{ij} \le \sigma_{ij}$) of
particles, $z_J = N_c/N'$, where $N'=N-N_r$, $N_r$ is the number of
rattler particles with fewer than three contacts, and $N_c$ only
includes overlapping pairs among the $N'$ particles within the `true'
contact network.  It is crucial to perform an error analysis on 
the contact number $z_J$, which is described in Appendix~\ref{error}.   

\begin{figure}
\scalebox{0.45}{\includegraphics{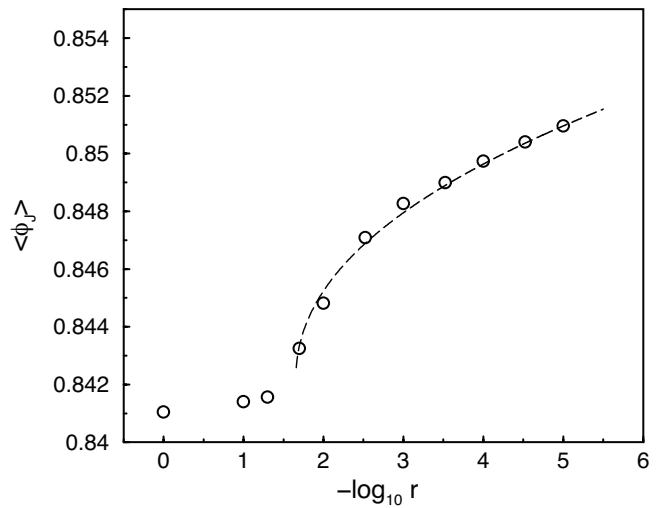}}
\caption{Average packing fraction $\langle \phi_J \rangle$ obtained
from Protocol $1$ as a function of the negative logarithm of the
thermal quench rate $r$ for $N=1024$. Data points at each rate
represent an average over typically $300$ static, amorphous
packings. The dashed line shows the scaling $\langle \phi_J \rangle
\sim [\log_{10} (r - r^*)]^{\mu}$, where $\mu \sim 0.5$ and $r^*
\approx 0.03$ is the thermal quench rate above which $\langle \phi_J
\rangle \approx 0.841$ is independent of $r$. }
\label{rate}
\end{figure}

\paragraph*{Packing Fraction}
We show results for the average packing fraction $\langle \phi_J
\rangle$ versus thermal quench rate $r$ over five orders of magnitude
obtained from Protocol $1$ in Fig.~\ref{rate}.  For large rates $r >
r^* \approx 0.03$, the average packing fraction $\langle \phi_J
\rangle \rightarrow 0.841$ is independent of rate, which agrees with
studies that employ athermal compression/decompression
packing-generation algorithms~\cite{longJ,ning}.  For $r < r^*$,
$\langle \phi_J \rangle$ increases approximately as $[\log_{10}
(r-r^*)]^{0.5}$ with decreasing rate.  We emphasize that all packings
used to present the data in Fig.~\ref{rate} are amorphous and
isostatic. Since $\langle \phi_J \rangle$ increases so slowly, it is
not possible to approach $\phi_{\rm xtal}$ using protocol
$1$.  Using an extrapolation, we estimate that rates
below $10^{-45}$ are required to reach $\phi_{\rm xtal}$, and thus we
employed Protocol $2$, not $1$, to generate compositionally and
positionally ordered packings.
  
\begin{figure}
\scalebox{0.45}{\includegraphics{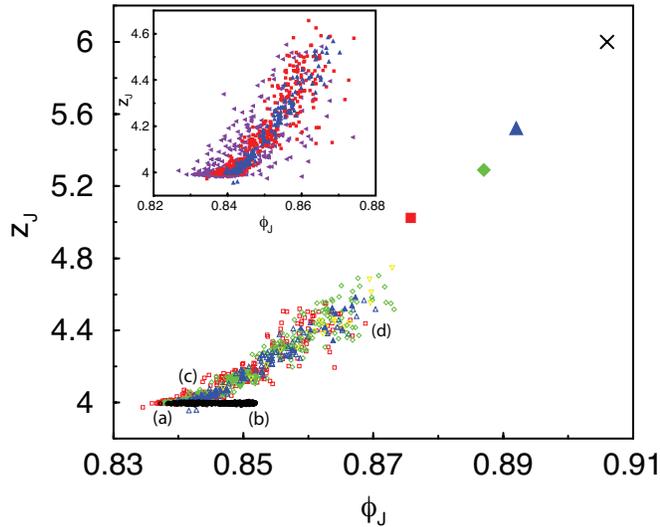}}
\caption{Scatter plot of the contact number $z_J$ versus the packing
fraction at jamming onset $\phi_J$.  The open circles indicate static
packings that were generated using Protocol $1$ for $N=1024$, while
all other symbols indicate static packings generated using Protocol
$2$.  The open squares, diamonds, and triangles correspond to
$N=1024$, $2048$, and $4096$, respectively, for all partitions $s$ and
systems with two partitions and random particle placements. The filled
squares, diamonds, upward triangles, and downward triangles correspond
to $N=1024$, $2048$, $4096$, and $8192$, respectively, for the systems with
two partitions and initial crystal lattice positions.  The black cross
indicates the values $z_J=6$ and $\phi_J=\pi/2\sqrt{3}$ for the
hexagonal crystal.  The labels (a)-(d) correspond to the images in
Fig.~\ref{picture}.  The inset shows the system-size dependence for
systems with two partitions and random initial positions at $N=256$
(leftward triangles), $1024$ (squares), and $4096$ (upward
triangles).}
\label{scatter}
\end{figure}

\begin{figure}
\scalebox{0.4}{\includegraphics{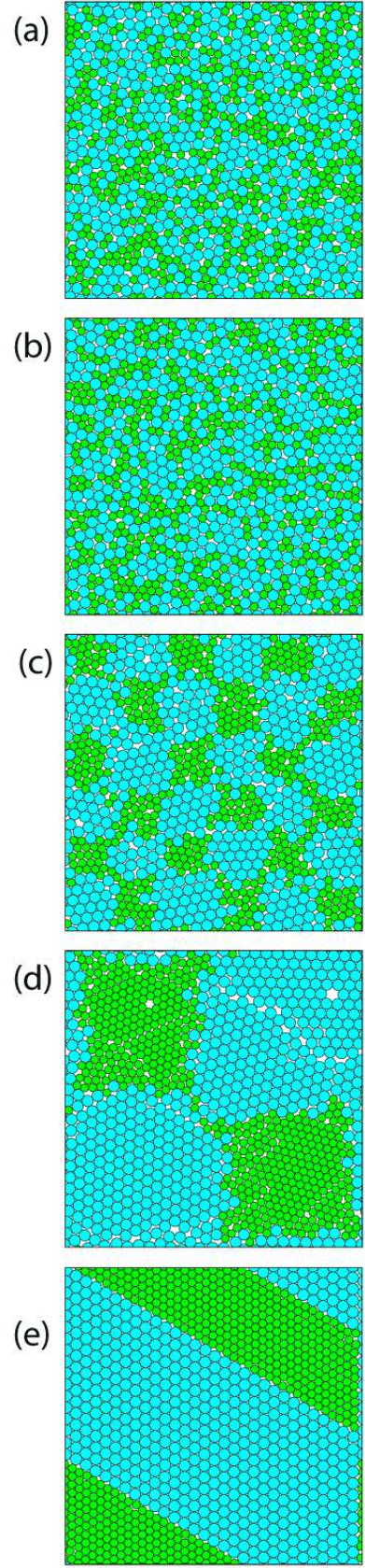}}
\caption{Images of representative static packings from the scatter
plot in Fig.~\ref{scatter} with (a) $\phi_J=0.837$, $z_J=3.99$, (b)
$\phi_J=0.853$, $z_J=4.00$, (c) $\phi_J=0.846$, $z_J=4.04$, (d)
$\phi_J=0.860$, $z_J=4.41$, and (e) $\phi_J=0.892$, $z_J \simeq 4.1$. (See 
Appendix~\ref{error}.)}
\label{picture}
\end{figure}

\paragraph*{Contact Number}
In Fig.~\ref{scatter}, we display a scatter plot of the contact number
$z_J$ versus $\phi_J$ for all static packings (where the contact
number is insensitive to the definition of `contact') generated using
Protocols $1$ and $2$.  (See Appendix~\ref{error} for a discussion of
the sensitivity of the contact number on the definition of contacting
particles.)  Fig.~\ref{scatter} shows several compelling
features. First, nearly all of the static packings obtained from
Protocol $1$ (open circles) are isostatic with $z_J = 4$, but they
occur over a range of packing fractions $\phi_{\rm min} \le \phi_J \le
\phi_{\rm max}$, where $\phi_{\rm min} = 0.837$ and $\phi_{\rm max} =
0.853$.  As shown in Appendix~\ref{error} $\phi_{\rm max}$ is likely
only a lower bound for the largest packing fraction at which isostatic
packings can occur in these systems.  Second, we find a cluster of
data points for Protocol $2$, for which the average $z_J$ is strongly
correlated---varying roughly linearly---with $\phi_J$.  The cluster
originates near $\phi_J \approx 0.84$, $z_J = z_{\rm iso} = 4$.  In
the inset to Fig.~\ref{scatter}, we show that the width of the cluster
of data points from Protocol $2$ narrows with increasing system size,
but the approximate linear relationship between the average $z_J$ and
$\phi_J$ is maintained.  Images of five representative packings from
the scatter plot in Fig.~\ref{scatter} are displayed in
Fig.~\ref{picture}.

\begin{figure}
\scalebox{0.4}{\includegraphics{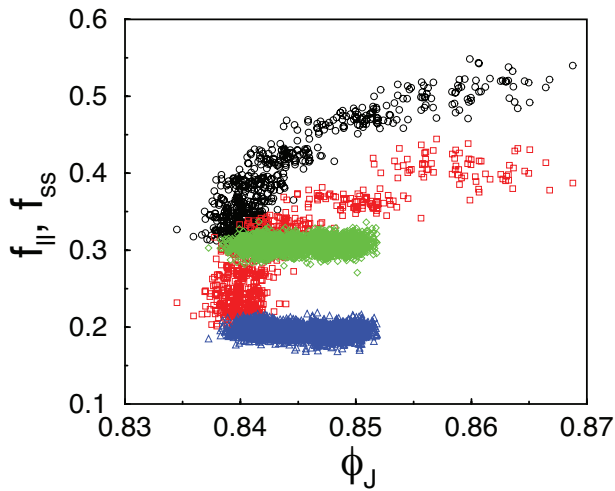}}
\caption{Scatter plot of the fraction of contacts between two large
$f_{ll}$ or two small particles $f_{ss}$ versus packing fraction
$\phi_J$ for all static packings from both protocols. The diamonds
(circles) and triangles (squares) display data from Protocol $1$ ($2$)
for $f_{ll}$ and $f_{ss}$, respectively.}
\label{ss}
\end{figure}

\paragraph*{Compositional Order}
We now describe measurements of the compositional and positional order
for static packings.  For the compositional order, we quantify the
fraction of overlapping pairs ($r_{ij} \le \sigma_{ij}$) that involve
two small $f_{ss}$ or large $f_{ll}$ particles.  A scatter plot of
$f_{ll}$ and $f_{ss}$ versus $\phi_J$ for static packings generated
from both protocols is shown in Fig.~\ref{ss}.  The packings from
Protocol $1$ show no signs of phase separation with $f_{ss} + f_{ll}
\approx f_{sl} \approx 0.5$ for all packings.  In contrast, Protocol
$2$ generates static packings with a range of compositional order as shown in
Fig.~\ref{picture} (c)-(e).  For example, at the largest $\phi_J$,
the system displays macrophase separation with $f_{ss} + f_{ll} \approx 1$ and
$f_{sl} \approx 0$.  We find similar results when we define contacting
pairs as those with $r_{ij} \le r_{\rm min} \sigma_{ij}$, where
$r_{\rm min}$ is set by the first minimum in $g(r)$.

\begin{figure}
\scalebox{0.8}{\includegraphics{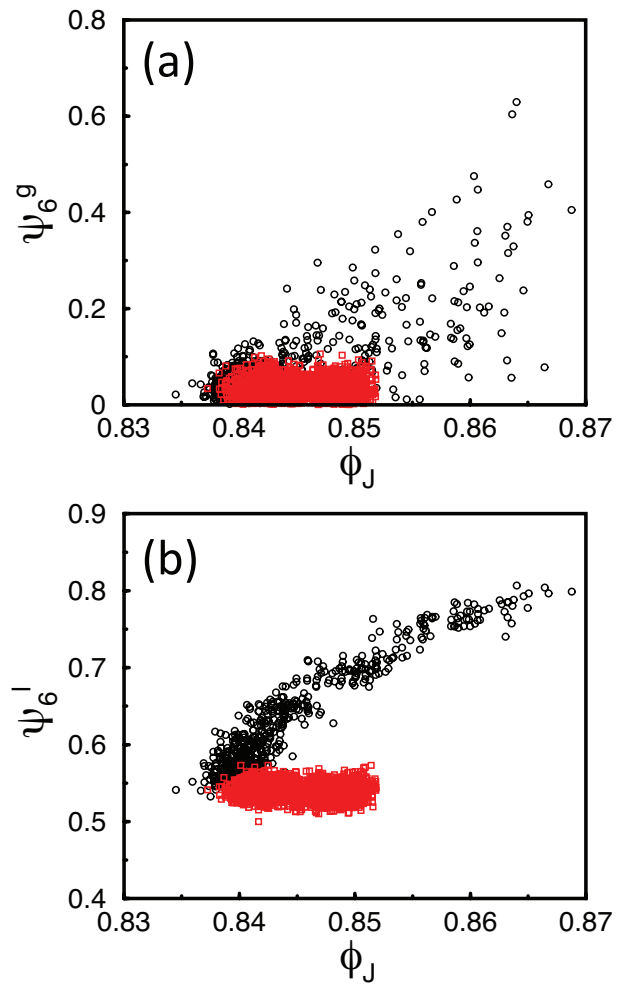}}
\caption{Scatter plot of the (a) global and (b) local bond
orientational order parameters, $\psi_6^g$ and $\psi_6^l$, versus 
packing fraction for static packings from protocol $1$ (squares) 
and $2$ (circles).}
\label{Q6}
\end{figure}

\paragraph*{Bond Orientational Order}
To quantify positional order, we calculate the bond orientational
order parameter $\psi_6$, which measures the hexagonal registry of nearest
neighbors~\cite{stein}. $\psi_6$ can be calculated `locally', which does
not consider phase information, or `globally', which allows phase
cancellations. A polycrystal will yield a relatively large value for the local
bond orientational order parameter $\psi_6^l$, even though the global
order parameter $\psi_6^g \sim 1/\sqrt{N_d}$, where $N_d$ is the number
of polycrystalline domains.  Eqs.~(\ref{2dglobal}) (global) and
(\ref{2dlocal}) (local) provide expressions for the bond orientational
order parameters in 2D.
\begin{eqnarray}
\label{2dglobal}
\psi_6^{g}&=&\frac{1}{N}\left|\displaystyle\sum_{i=1}^N\frac{1}{n_i}
\displaystyle\sum_{j=1}^{n_i}e^{6\imath\theta_{ij}}\right| \\
\label{2dlocal}
\psi_6^{l}&=&\frac{1}{N} \displaystyle\sum_{i=1}^N\frac{1}{n_i}\left|
\displaystyle\sum_{j=1}^{n_i}e^{6\imath\theta_{ij}}\right|, 
\end{eqnarray}
where $\theta_{ij}$ is the angle between a central particle $i$ and
neighbors $j$ and $n_i$ denotes the number of nearest neighbors of
$i$. Two particles are deemed nearest neighbors if their
center-to-center separation $r_{ij} < r_{\min} \sigma_{ij}$. 

\begin{figure}
\scalebox{0.45}{\includegraphics{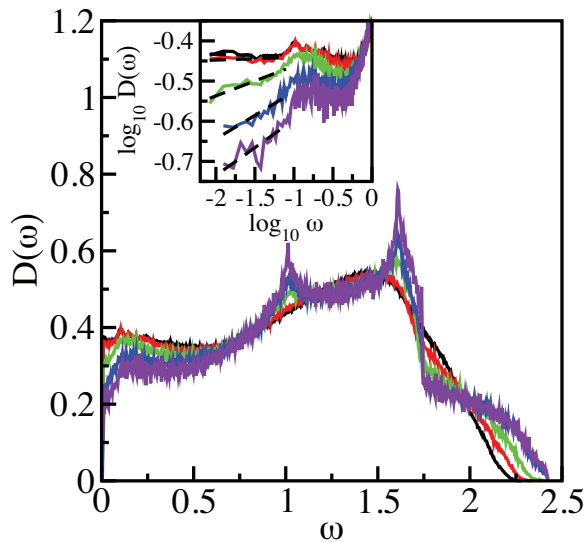}}
\caption{Density $D(\omega)$ of normal mode frequencies $\omega$ for
$N=1024$ bidisperse frictionless disk packings obtained using
Protocols $1$ and $2$ as a function of the contact number at jamming
onset for $z_J \simeq 4.0$ (black), $4.0 \le z_J \le 4.1$ (red), $4.1
\le z_J \le 4.2$ (green), $4.3 \le z_J \le 4.4$ (blue), and $4.5 \le
z_J \le 4.6$ (violet). The inset shows the same data except that it
focuses on low frequencies $\omega < 1$ and includes power-law fits to
$D(\omega) \sim \omega^{\alpha}$ as dashed lines.}
\label{DOS_bi}
\end{figure}

\begin{figure}
\scalebox{0.45}{\includegraphics{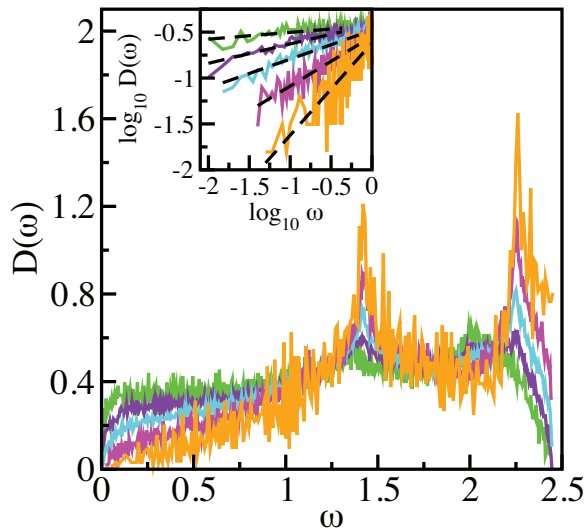}}
\caption{Density $D(\omega)$ of normal mode frequencies $\omega$ for
$N=1024$ {\it monodisperse} frictionless disk packings obtained using
Protocol $1$ as a function of the contact number at jamming onset for
$4.1 \le z_J \le 4.2$ (green), $4.5 \le z_J \le 4.6$ (violet), $4.9
\le z_J \le 5.0$ (cyan), $5.4 \le z_J \le 5.5$ (magenta), and $z_J
\simeq 6.0$ (orange). The inset shows the same data except that it
focuses on low frequencies $\omega < 1$ and includes power-law fits to
$D(\omega) \sim \omega^{\alpha}$ as dashed lines.}
\label{DOS}
\end{figure}

The results for the global and local bond orientational parameters
$\psi_6^g$ and $\psi_6^l$ are shown in Fig.~\ref{Q6}.  The static
packings obtained from Protocol $1$ possess only local bond
orientational order with $\psi_6^l \approx 0.55$ as found in dense
liquids~\cite{stein}, and $\psi_6^g \sim 1/\sqrt{N}$. Further, there is
no correlation between the packing fraction $\phi_J$ and global or
local bond orientational order.  In contrast, for the phase-separated
and partially crystalline packings from Protocol $2$, we find that
there is a strong positive correlation between $\psi_6^l$ and $\phi_J$
and a somewhat weaker correlation between $\psi_6^g$ and $\phi_J$.
 
The static packings from Protocols $1$ and $2$ have different
structural properties.  Those from $1$ are amorphous and possess
similar structural properties even though they exist over a range of
packing fraction.  In contrast, there is a positive correlation
between compositional and positional order and packing fraction for
the phase-separated and partially crystalline packings from Protocol
$2$.  We will now describe the mechanical properties of the static 
packings including the spectrum of normal modes and static shear modulus 
as a function of contact number and order.

\paragraph*{Spectrum of Normal Modes}
The spectrum of normal modes provides significant insight into the
structural and mechanical properties of mechanically stable
packings~\cite{longJ}.  For example, there is evidence that the
low-frequency region of the spectrum controls the
static shear response of jammed packings~\cite{ellipse}. To calculate the
spectrum, we diagonalize the dynamical matrix of all possible second
derivatives with respect to particle positions evaluated at positions
of the static packing---assuming that no existing contacts break and
no new contacts form~\cite{chapter}.  This yields $2N'-2$ nontrivial
eigenvalues $e_i$ after accounting for translational invariance.  We
consider here only mechanically stable packings, and thus all $2N' -
2$ of the eigenvalues are nonzero~\cite{foot}.

\begin{figure}
\scalebox{0.4}{\includegraphics{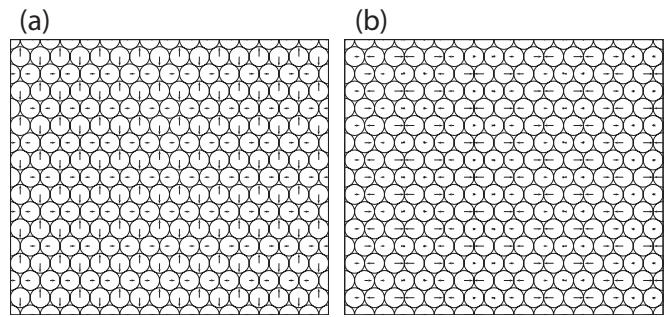}}
\caption{Eigenvectors corresponding to the modes with frequencies near
the (a) first and (b) second peaks in the density of states
$D(\omega)$ for monodisperse packings with $z_J \simeq 6$ and $\phi_J
\simeq \phi_{\rm xtal}$ for $N=256$.  The size of the eigenvector
component for each particle is proportional to the length of the vector
associated with each particle.}
\label{eigenvector}
\end{figure}

\begin{figure}
\scalebox{0.45}{\includegraphics{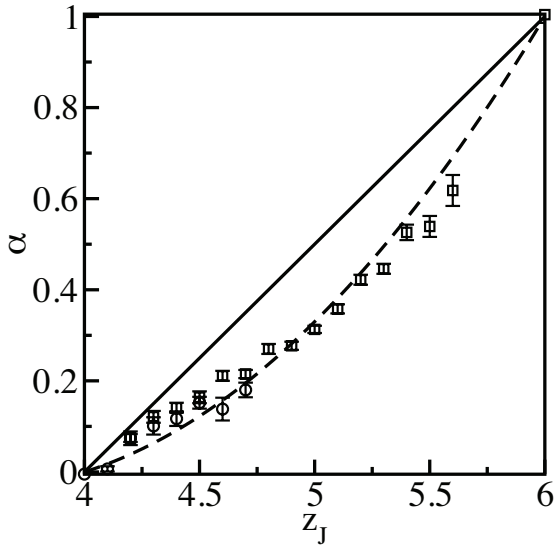}}
\caption{Power-law exponent $\alpha$ for the scaling of the density of
states with frequency in the limit $\omega \rightarrow 0$ ($D(\omega)
\sim \omega^{\alpha}$) as a function of contact number at jamming
onset $z_J$ for bidisperse (circles) and monodisperse (squares)
packings.  (The error bars indicate the error in $\alpha$ from
least-squares analysis.)  The dashed line is a fit to Eq.~\ref{alpha}
(with $a=0.17$), which interpolates the data between the limiting
values $\alpha = 0$ at $z_J = z_{\rm iso}=4$ and $\alpha=1$ (Debye
behavior) at $z_J=z_{\rm xtal}=6$. The solid line is Eq.~\ref{alpha}
with $a=0$.}
\label{exponent}
\end{figure}

The density $D(\omega)$ of normal mode frequencies $\omega_i =
\sqrt{e_i/N}$, or density of states (DOS), is given by $D(\omega) =
(N(\omega+\delta \omega)-N(\omega))/\delta \omega$, where $N(\omega)$
is the number of modes with frequency less than or equal to $\omega$.
The density of states $D(\omega)$ for packings of bidisperse
frictionless disks is shown in Fig.~\ref{DOS_bi} as a function of the
contact number at jamming onset $z_J$.  As in previous studies
\cite{longJ}, we find that for isostatic systems with $z_J \simeq 4$,
$D(\omega)$ possesses a nearly constant regime at low frequencies,
which signals an abundance of low-frequency modes compared to ideal
Debye behavior (where $D(\omega) \sim \omega$ as $\omega \rightarrow
0$) for ideal 2D harmonic solids.  For the micro- and macro-phase
separated bidisperse packings generated using Protocol $2$ with $z_J
\gtrsim 4.1$, the density of states develops two other interesting
features.  First, $D(\omega)$ develops two strong peaks near $\omega
\simeq 1.0$ and $1.6$ instead of a single broad peak centered near
$\omega \approx 1.4$ for isostatic amorphous systems.  (We will see 
below that these peaks are associated with crystallization.)  Second, we
observe that as $z_J$ increases and the packings become hyperstatic,
the weight in $D(\omega)$ at low frequency ($\omega \lesssim 0.3$)
decreases.  As shown in the inset to Fig.~\ref{DOS_bi}, the density of
states scales as a power-law
\begin{equation}
\label{dos}
D(\omega) \sim \omega^{\alpha}
\end{equation}
in the limit $\omega \rightarrow 0$ with a scaling exponent $\alpha$
that varies continuously with contact number $z_J$ as shown in
Fig.~\ref{exponent}.  (See Appendix~\ref{system_size} for a discussion
of the system-size dependence of the exponent $\alpha$.)  Note,
however, that the plateau in the density of states remains largely
unchanged in the intermediate frequency regime $0.3 \le \omega
\lesssim 1$ over a wide range of $z_J$, which implies that some of the
remarkable features of jamming in isostatic systems also hold for
hyperstatic systems.

\begin{figure}
\scalebox{0.45}{\includegraphics{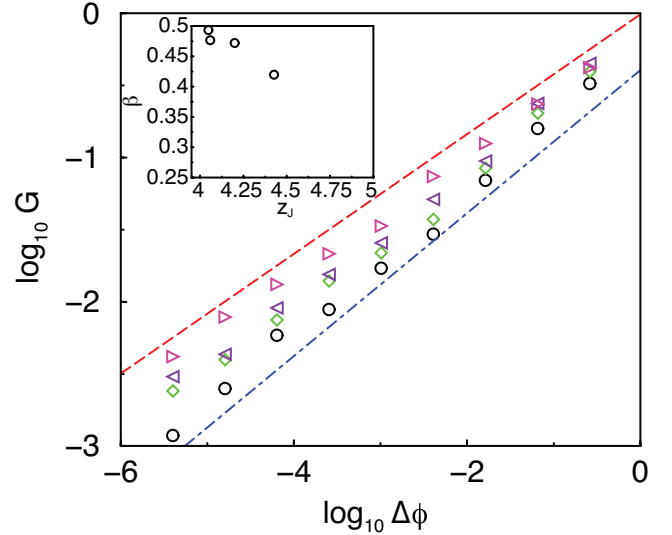}}
\caption{Static shear modulus $G$ versus the deviation in packing
fraction from the jamming onset $\Delta \phi = \phi-\phi_J$ for static
packings at $\langle z_J\rangle = 4.0$ (circles), $4.15$ (diamonds),
$4.35$ (left triangles), and $4.55$ (right triangles).  The long dashed
(dot-dashed) line has slope $0.5$ ($0.4$).  The inset shows the 
power-law scaling exponent $\beta$ for the static shear
modulus ($G \sim (\Delta \phi)^{\beta}$) versus the contact number
$z_J$ at jamming.}
\label{G}
\end{figure}

To test the generality of the results for the density of states, we
also calculated $D(\omega)$ for monodisperse frictionless disk
packings generated using Protocol $1$ as shown in Fig.~\ref{DOS}.  The
density of states for monodisperse systems displays similar features
to that for bidisperse systems. 1. A plateau in $D(\omega)$ exists at
low to intermediate frequencies for nearly isostatic systems.
2. Strong distinct peaks are located near $\omega \simeq 1.4$ and
$2.25$ for hyperstatic packings.  Eigenvectors that correspond to the
two peak frequencies are visualized in Fig.~\ref{eigenvector}. 3. A
power-law regime $D(\omega) \sim \omega^{\alpha}$ develops in the
$\omega \rightarrow 0$ limit for hyperstatic packings.  The exponent
$\alpha$ varies continuously with $z_J$ with a similar functional
dependence to that for bidisperse systems as shown in
Fig.~\ref{exponent}.  A notable difference between bidisperse and
monodisperse systems is that a continuous power-law regime in
$D(\omega)$ persists to higher frequencies ($\omega \sim 1$) for
monodisperse compared to bidisperse systems.

The dependence of the scaling exponent $\alpha$ on $z_J$ is displayed
for all bidisperse and monodisperse packings (binned by $z_J$) in
Fig.~\ref{exponent}.  We find that $\alpha$ increases monotonically
with $z_J$ and use the suggestive empirical form
\begin{equation}
\label{alpha}
\alpha = (d-1)\frac{z_J-z_{\rm iso}}{z_{\rm xtal}-z_{\rm iso}} + 
a (z_J-z_{\rm iso})(z_J-z_{\rm xtal}),
\end{equation}
where $a$ is a fitting parameter, to describe the data between the
limiting values $\alpha=0$ at $z_J = z_{\rm iso}$ and $\alpha = d-1$
(Debye behavior) at $z_J = z_{\rm xtal}$.  The continuous increase in
$\alpha$ from $0$ to $1$ as the contact number increases suggests a
different scenario for the behavior of the jamming transition as a
function of $z_J$ and positional order compared to the
first-order-like transition found as the system compacts above random
close packing in simulations of frictional granular
materials~\cite{makse}.  

\paragraph*{Static Shear Modulus}
To measure the static linear shear modulus $G$, we slightly deform the
system by applying an infinitesimal simple shear strain $\gamma$
(along the $x$-direction with gradient in the $y$-direction), allowing
the system to relax via energy minimization at fixed strain, and then
measuring the resulting shear stress response, $G=
d\Sigma_{xy}/d\gamma$.  In Fig.~\ref{G}, we show the shear modulus
versus the amount of compression $\Delta \phi = \phi - \phi_J$ for
bidisperse packings obtained from Protocols $1$ and $2$ at several
values of $z_J$.  We find generally that in the limit $\Delta \phi
\rightarrow 0$ the static shear modulus scales as a power-law with
$\Delta \phi$:
\begin{equation}
\label{geq}
G = G_0 (\Delta \phi)^{\beta},
\end{equation}
where the scaling exponent $\beta$ (and prefactor $G_0$) depend on
$z_J$.  As shown in Fig.~\ref{G}, $\beta$ decreases steadily from
$0.5$ to $0.4$ as the contact number $z_J$ at jamming increases.  Note
that $\beta=0.5$ for $z_J = z_{\rm iso}$ was obtained in previous work
on isostatic packings~\cite{longJ}.  The results in Fig.~\ref{G}
suggest that the critical behavior ({\it e.g.}  power-law scaling of
the shear modulus) found in jammed isostatic systems persists when the
jamming onset is hyperstatic.  Further studies are required to
determine whether the scaling exponent for the static shear modulus can be
varied over the full range from $0.5$ to $0$.

\begin{figure}
\scalebox{0.4}{\includegraphics{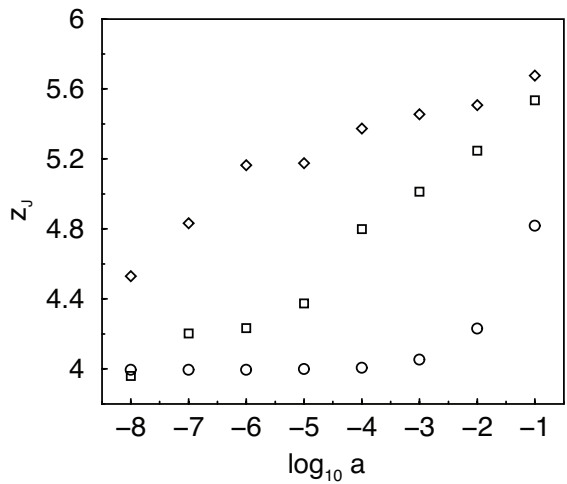}}
\caption{The contact number $z_J$ as a function of $a$, where the
condition $r_{ij} \le (1+a)\sigma_{ij}$ determines whether particles
$i$ and $j$ are in contact. The packings shown are $N=1024$, $\phi_J =
0.837$ (circles); $N=1014$, $\phi_J=0.892$ (squares); and $N=2390$,
$\phi_J=0.897$ (diamonds).}
\label{loga}
\end{figure}

\section{Conclusions}
\label{conclusions}

Using computer simulations, we generated a large library of
mechanically stable packings of bidisperse, frictionless disks that
span a wide range of contact number from $z_J=z_{\rm iso}=4$ to
$z_{\rm xtal}=6$ and packing fraction at jamming from $\phi_J \sim
0.84$ to near $\phi_{\rm xtal}$.  We find that there is an amorphous,
isostatic branch of packings that spans a finite range in packing
fraction in the large-system limit.  Over this range of packing
fraction, these packings are amorphous with no correlation between
bond orientational order or compositional order and $\phi_J$.  We also
find a branch of phase-separated and partially crystalline packings
for which the compositional and positional order increase with
$\phi_J$.  In addition, we characterize the mechanical properties of
the static packings by measuring the spectrum of normal modes and the
static shear modulus.  We find that the mechanical properties of the
packings vary {\it continuously} as the contact number and structural
and compositional order at jamming onset increase from their isostatic
values.  In particular, we find that the static shear modulus scales
as a power-law in the amount of compression, $G\sim (\Delta
\phi)^{\beta}$, and that the low-frequency density of states scales as
a power-law in frequency, $D(\omega) \sim \omega^{\alpha}$, and both
$\alpha$ and $\beta$ vary continuously with contact number at jamming
onset.  These findings emphasize that jamming behavior in systems with
purely repulsive contact potentials occurs over a range of contact
numbers, not just near $z_J = z_{\rm
iso}$~\cite{hatano,hatano2,otsuki}.  In future studies, we will
investigate the relationship between the scaling exponents $\alpha$
and $\beta$, which is likely an important feature of jamming in
hyperstatic systems.

\begin{figure}
\scalebox{0.4}{\includegraphics{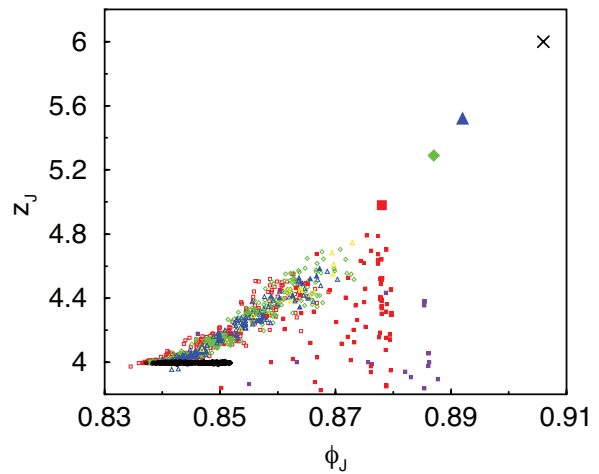}}
\caption{Contact number $z_J$ versus packing fraction $\phi_J$ for the
same data in Fig.~\ref{scatter} and an additional set of packings
obtained from thermalizing the configurations in Fig.~\ref{scatter}
with $\phi_J > 0.86$ and then identifying the nearest packing.  The 
variation in $z_J$ increases with $\phi_J$.}
\label{all}
\end{figure}
      
\section{Acknowledgments}
We thank the organizers of the Frontiers in Nonequilibrium Physics and
YKIS2009 workshops.  We also acknowledge A. Donev, R. Hoy, and
M. Shattuck for helpful conversations. This research was supported by
the National Science Foundation under Grant Nos. CBET-0828359 (LS),
DMS-0835742 (CO, CS), and PHY-0551164.  We thank the Kavali Institute 
for Theoretical Physics for their hospitality during ``The Physics of 
Glasses: Relating Metallic Glasses to Molecular, Polymeric and 
Oxide Glasses'' Program.   This work also benefited from the
facilities and staff of the Yale University Faculty of Arts and
Sciences High Performance Computing Center and NSF grant
no. CNS-0821132 that partially funded acquisition of the computational
facilities.

\appendix

\section{Error analysis of contact number}
\label{error}

In this appendix, we study how sensitive the contact number $z_J$ is
to the definition of whether two particles are in contact.  In
Fig.~\ref{loga}, we show $z_J$ versus $\log_{10} a$ where two disks
$i$ and $j$ are considered in contact (or overlapping) if $r_{ij} \le
(1+a) \sigma_{ij}$ for three representative configurations: $N=1024$,
$\phi_J = 0.837$ (circles); $N=1014$, $\phi_J=0.892$ (squares); and
$N=2390$, $\phi_J=0.897$ (diamonds).  We see that the contact number
is well-defined for amorphous configurations at low packing fractions,
{\it i.e.} the contact number is constant over a wide range of $a$
that determines whether two particles are in contact.  In contrast,
for packings with large $\phi_J$ and significant order as shown in 
Fig.~\ref{picture} (e), the contact
number varies continuously with $a$ down to the numerical precision of
the particle positions in the simulations ($a_{\rm min} \sim
10^{-8}$).  Thus, at the current numerical precision of the
simulations it is difficult to determine $z_J$ accurately for the
partially ordered and ordered configurations.  To test the robustness
of the contact numbers, we also added weak thermal fluctuations to the
packings with $\phi_J > 0.855$ in Fig.~\ref{scatter} for times
significantly shorter than the structural relaxation time, and then
found the nearest static packing.  This data, shown by the small
filled symbols in Fig.~\ref{all}, possess surprisingly small contact
numbers and begin to fill in the region at large $\phi_J$ and small
$z_J$.  As a result, we only include configurations in
Fig.~\ref{scatter} that possess plateaus in $z_J$ versus $a$ over a
range $a_{\rm min} \le a \le a_{\rm max}$ of at least two orders of
magnitude.
      
\section{Robustness of the Density of States}
\label{system_size}

In this appendix, we test the robustness of our measurements of the
the density of states $D(\omega)$ by (1) studying the system-size
dependence of the accumulated frequency distribution $N(\omega)$ and
(2) comparing $D(\omega)$ for hyperstatic packings at jamming onset
with contact number $z_J$ to that for overcompressed packings at the 
same contact number $z=z_J$.

\begin{figure}
\scalebox{0.4}{\includegraphics{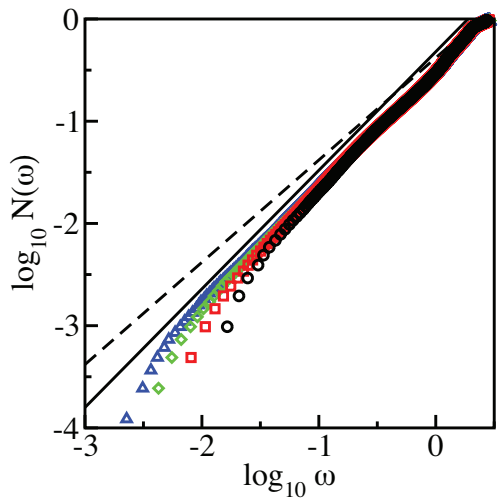}}
\caption{Number $N(\omega)$ of normal modes of the dynamical matrix
with frequency less than or equal to $\omega$ for bidisperse packings
at jamming onset generated using Protocol $2$ with $4.4 \le z_J \le
4.5$ and $N=512$ (circles), $1024$ (squares), $2048$ (diamonds), and
$4096$ (triangles).  The solid (dashed) line has slope $1.16$ ($1$).}
\label{abi}
\end{figure}

To eliminate noise from numerical differentiation, we calculate the
accumulated distribution $N(\omega) = \int_0^{\omega} D(\omega')
d\omega'$ (number of modes with frequency less than or equal to
$\omega$).  For reference, we first show $N(\omega)$ for monodisperse
packings at jamming onset with $z_J \simeq 6$ and $\phi_J \simeq
\phi_{\rm xtal}$ as a function of system size for $N=16$ to $6400$.
The crystalline systems show robust Debye power-law scaling $N(\omega)
\sim \omega^2$ at low frequency for all system sizes.  $N(\omega)$ for
bidisperse packings at jamming onset is shown in Fig.~\ref{abi} for
$4.4 \le z_J \le 4.5$ as a function of system size.  $N(\omega)$
displays a power-law scaling with an exponent that approaches $1+\alpha
= 1.16 > 1$ in the large-system limit.  Similar robust scaling
exponents are found for all $z_J$.
 
\begin{figure}
\scalebox{0.4}{\includegraphics{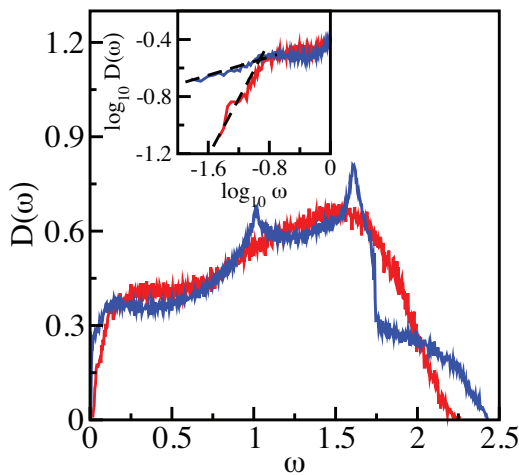}}
\caption{The density of normal modes $D(\omega)$ with frequency
$\omega$ for bidisperse packings at jamming onset generated using
Protocol $2$ with $4.4 \le z_J \le 4.5$ (blue line) and overcompressed
packings with contact number $z$ in the same range (red line).  The dashed
lines in the inset have slope $0.16$ and $1$.}
\label{compress}
\end{figure}

Distinctive features of the density of states $D(\omega)$ for
hyperstatic bidisperse packings at jamming onset are the power-law
scaling of $D(\omega) \sim \omega^{\alpha}$ at the lowest frequencies,
where $\alpha$ varies continuously with $z_J$, and the persistence of
the plateau in $D(\omega)$ at intermediate frequencies over a range of
$z_J$.  Do highly compressed packings display these same features?  In
Fig.~\ref{compress}, we compare $D(\omega)$ for hyperstatic packings
at jamming onset with $4.4 \le z_J \le 4.5$ and overcompressed
packings in the same range of contact number $z \sim z_J$.  For the
overcompressed packings, we find that $D(\omega) \sim \omega^{\alpha}$,
with $\alpha=1$, while $\alpha \approx 0.16$ at the lowest frequencies
with a crossover to a plateau at intermediate frequencies for the
hyperstatic packings at jamming onset.  Thus, hyperstatic packings at
jamming onset possess significantly more low-frequency normal modes
than overcompressed systems at the same contact number as shown in the
inset to Fig.~\ref{compress}.

\begin{figure}
\vspace{0.3in}
\scalebox{0.4}{\includegraphics{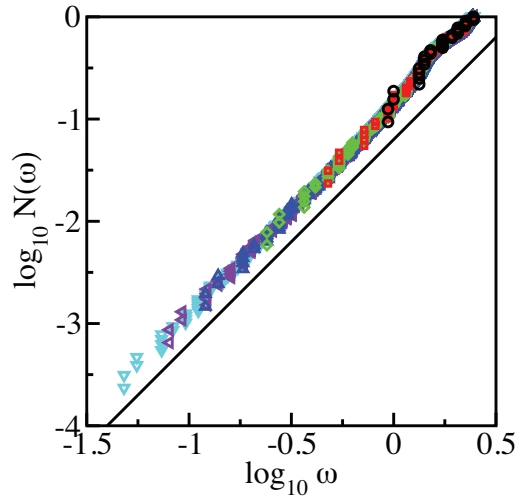}}
\caption{Number $N(\omega)$ of normal modes of the dynamical matrix
with frequency less than or equal to $\omega$ for monodisperse
packings at jamming onset with $z_J \simeq 6$ and $\phi_J \simeq \phi_{\rm
xtal}$ and $N=16$ (circles), $64$ (squares), $256$ (diamonds), $1024$
(upward triangles), $2304$ (leftward triangles), and $6400$ (downward
triangles).  The solid line has slope $2$.}
\label{accumulate}
\end{figure}

\begin{references}
\bibitem{foot4} We assume that the largest packing fraction in
bidisperse systems is the macrophase-separated packing with $\phi_J
\approx \phi_{\rm xtal}$ in the large-system limit.
\bibitem{foot} We use a threshold of $e_{\rm min}=10^{-6}$ above
which eigenvalues are deemed nonzero.
\end{references}
\end{document}